\documentclass[aps,pre,twocolumn,superscriptaddress]{revtex4}

\usepackage{graphicx}
\usepackage{amssymb}
\usepackage{epstopdf}
\usepackage{subfigure}

\begin{document}
%\preprint{}

%Title of paper
\title{A Supersymmetric Approach to  Excited States via Quantum Monte Carlo}
\author{Eric R. Bittner}
\email[email:]{bittner@uh.edu}
\affiliation{Dept. of Chemistry, University of Houston, Houston TX 77204}

\author{Jeremy B. Maddox}
\affiliation{
Department of Chemistry and Biochemistry,
Texas Tech University 
Lubbock, TX 79409-1061}

\author{Donald J. Kouri}
\email[email:]{kouri@uh.edu}
\affiliation{Dept. of Chemistry, University of Houston, Houston TX 77204}

\begin{abstract}
We present here a supersymmetric (SUSY) approach for determining
excitation energies within the context of a quantum Monte Carlo
scheme.  By using the fact that SUSY quantum mechanics gives rises to
a series of isospectral Hamiltonians, we show that Monte Carlo
ground-state calculations in the SUSY partners can be used to
reconstruct accurately both the spectrum and states of an arbitrary
Schr\"odinger equation.  Since the ground-state of each partner
potential is node-less, we avoid any ``node''-problem typically
associated with the Monte Carlo technique.  While we provide an
example of using this approach to determine the tunneling states in a
double-well potential, the method is applicable to any 1D potential
problem.  We conclude by discussing the extension to higher
dimensions.
\end{abstract}


%\keywords{}

\maketitle

\section{Introduction}

The variational Monte Carlo (VMC) technique is a powerful way to
estimate the ground state of a quantum mechanical system.  The basic
idea is that one can use the variational principle to minimize energy
expectation value with respect to a set of parameters $\{\alpha\}$
\begin{eqnarray}
E(\alpha)
&=&
\frac{\int |\psi(x,\alpha)|^{2}(H\psi)/\psi(x,\alpha)) dx}{\int |\psi(x,\alpha)|^{2}dx}.
\end{eqnarray}
Following the Monte Carlo method for evaluating integrals, one intreprets
\begin{eqnarray}
p(x)dx = \frac{|\psi(x,\alpha)|^{2}dx}{\int |\psi(x,\alpha)|^{2}dx}
\end{eqnarray}
as a probability distribution function.  Typically, one assumes a
functional form for the trial wave function, $\psi(x,\alpha)$ and the
numerical advantage is that one can evaluate the energy integral by
simply evaluating $\psi(x,\alpha)$.  The method becomes variational
when one then adjusts the parameters to optimize the trial wave
function.  Since the spectrum of $H$ is bounded from below, the
optimized trial wave-function provides a best approximation to the
true ground state of the system.  However, since $p(x) =
|\psi(x,\alpha)^{2}$ is a positive definite function, this procedure
fails if the system has nodes or if the position of the nodes is
determined by the parameters.  One can in principle obtain excitation
energies by constraining the trial function to have a fixed set of
nodes perhaps determined by symmetry.

Given that (VMC) is a robust technique for ground states, it would be
highly desirable if the technique could be extended to facilitate the
calculation of excited states.  In this paper, we present such an
extension (albeit in one dimension) using supersymmetric (SUSY)
quantum mechanics.  The underlying mathematical idea behind SUSY is
that every Hamiltonian $H_{1} = T + V_{1}$ has a partner Hamiltonian,
$H_{2} = T + V_{2}$ ($T$ being the kinetic energy operator) in which
the spectrum of $H_{1}$ and $H_{2}$ are identical for all states above
the ground state of $H_{1}$.  That is to say, the ground state of
$H_{2}$ has the same energy as the first excited state of $H_{1}$ and
so on.  This hierarchy of related Hamiltonians and the algebra
associated with the SUSY operators present a powerful formal approach
to determine the energy spectra for a wide number of systems.
\cite{inomata:3638,gunther:54,balents:12970,cannata:209,rodrigues:125023,niederle:1280,berezinsky:034007,leung:4802,c.:236,humi:083515,margueron:024318}
To date, little has been done exploiting SUSY as a way to develop new
numerical techniques.

In this paper, we shall use the ideas of SUSY-QM to develop a Monte
Carlo-like scheme for computing the tunneling splittings in a
symmetric double-well potential.  While the model can be solved solved
using other techniques, this provides a useful proof of principle for
our approach.  We find that the the SUSY/VMC combination provides a
useful and accurate way to obtain the tunneling splitting for this
system and excited state wave function for this system.  While our
current focus is on a one-dimensional system, we conclude by
commenting upon how the technique can be extended to multi-particle
systems and to higher dimension.  In short, our results strongly
suggest that this approach can be brought to bear on a more general
class of problems involving multiple degrees of freedom.
Surprisingly, the connection between the Monte Carlo technique and the
SUSY hierarchy has not been exploited until this paper.

%Once we have determined the ground state of $H_{1}$, we use this to generate the partner potential, then re-apply the VMC approach
%to determine the ground state of $H_{2}$. In doing so we determine both the excitation energy and the 
%first excited state wavefunction  of $H_{1}$.    Furthermore, our variational approach 
%for optimizing the trial state (for either $H_{1}$ or its partner $H_{2}$) does not impose any 
%implicit symmetry on the final solution.  

\section{Supersymmetric Quantum Mechanics}
Supersymmetric quantum mechanics (SUSY-QM) is obtained by factoring
the Schr\"odinger equation into the form
\cite{Witten:1981bq,Witten:1982nx,Cooper:1995jt}
\begin{eqnarray}
H\psi = A^{\dagger}A \psi_{o}^{(1)} = 0
\end{eqnarray}
using the operators
\begin{eqnarray}
A  = \frac{\hbar}{\sqrt{2m}} \partial_{x}+ W \\
A^{\dagger}  = -\frac{\hbar}{\sqrt{2m}}\partial_{x} + W .
\end{eqnarray}
Since we can impose  $A\psi_{o}^{(1)} = 0$, we can immediately write that 
\begin{eqnarray}
W(x) = -\frac{\hbar}{\sqrt{2m}}\partial_x \ln\psi_{o}.
\end{eqnarray}
$W(x)$ is the {\em superpotential} which is related to the physical 
potential by a Riccati equation.
\begin{eqnarray}
V(x) = W^{2}(x)  - \frac{\hbar}{\sqrt{2m}}W'(x).
\end{eqnarray}
The SUSY factorization of the Schr\"odinger can always be applied in one-dimension.

From this point on we label the original Hamiltonian operator and its
associated potential, states, and energies as $H_{1}$, $V_{1}$,
$\psi_{n}^{(1)}$ and $E_{n}^{(1)}$.  One can also define a partner
Hamiltonian, $H_{2} = AA^{\dagger}$ with a corresponding potential
\begin{eqnarray}
V_{2} = W^{2} +  \frac{\hbar}{\sqrt{2m}}W'(x).
\end{eqnarray}
All of this seems rather circular until one recognizes that $V_{1}$
and its partner potential, $V_{2}$, give rise to a common set of
energy eigenvalues.  This principle result of SUSY can be seen by
first considering an arbitrary stationary solution of $H_{1}$,
\begin{eqnarray}
H_{1} \psi_{n}^{((1)} = A^{\dagger}A\psi_{n} = E_{n}^{(1)}\psi_{n}^{(1)}.
\end{eqnarray}
This implies that $(A\psi_{n}^{(1)})$ is an eigenstate of $H_{2}$ with
energy $E_{n}^{(1)}$ since
\begin{eqnarray}
H_{2}(A\psi_{n}^{(1)}) = AA^{\dagger}A\psi_{n}^{(1)} = E_{n}^{(1)}(A\psi_{n}^{(1)}).
\end{eqnarray}
Likewise, the Schr\"odinger equation involving the partner potential
$H_{2}\psi_{n}^{(2)} = E_{n}^{(2)}\psi_{n}^{(2)} $ implies that
\begin{eqnarray}
A^{\dagger}AA^{\dagger}\psi_{n}^{(2)} =
H_{1}(A^{\dagger}\psi_{n}^{(2)}) =
E_{n}^{(2)}(A^{\dagger}\psi_{n}^{(2)}).\label{chargeop}
\end{eqnarray}
This (along with $E_{o}^{(1)} = 0$ ) allows one to conclude that the
eigenenergies and eigenfunctions of $H_{1}$ and $H_{2}$ are related in
the following way: $ E_{n+1}^{(1)} = E_{n}^{(2)}, $
\begin{eqnarray}
\psi_{n}^{(2)} = \frac{1}{\sqrt{E_{n+1}^{(1)}}} A \psi_{n+1}^{(1)},\, \,{\rm and} \,\,
\psi_{n+1}^{(1)} = \frac{1}{\sqrt{E_{n}^{(2)}}} A^{\dagger} \psi_{n}^{(2)} 
\end{eqnarray}
for $n > 0$.  
\footnote{Our notation from here on is that $\psi_{n}^{(m)}$ denotes
  the $n$th state associated with the $m$th partner Hamiltonian with
  similar notion for related quantities such as energies and
  superpotentials. } Thus, the {\em ground state of $H_{2}$ has the
  same energy as the first excited state of $H_{1}$}.  If this state
$\psi_{o}^{(2)}$ is assumed to be node-less, then $\psi_{1}^{(1)}
\propto A^{\dagger}\psi_{o}^{(2)}$ will have a single node.  We can
repeat this analysis and show that $H_{2}$ is partnered with another
Hamiltonian, $H_{3}$ whose ground state is isoenergetic with the first
excited state of $H_{2}$ and thus isoenergetic with the second excited
state of the original $H_{1}$.  This hierarchy of partners persists
until all of the bound states of $H_{1}$ are exhausted.

\section{Adaptive Monte Carlo }
Having defined the basic terms of SUSY quantum mechanics, let us
presume that one can determine an accurate approximation to the ground
state density $\rho_{o}^{(1)}(x)$ of Hamiltonian $H_{1}$.  One can
then use this to determine the superpotential using the Riccati
transform
\begin{eqnarray}
W_{o}^{(1)} = -\frac{1}{2} \frac{\hbar}{\sqrt{2m}} \frac{\partial \ln\rho_{o}^{(1)}}{\partial x} \label{riccati}
\end{eqnarray}
and the partner potential
\begin{eqnarray}
V_{2} = V_{1} - \frac{\hbar^{2}}{2m} \frac{\partial^{2} \ln\rho_{o}^{(1)}}{\partial x^{2}}. \label{eq13}
\end{eqnarray}
Certainly, our ability to compute the energy of the ground state of the partner 
potential $V_{2}$ depends on having first obtained an accurate estimate of the 
ground-state density associated with the original $V_{1}$.   

For this we turn to an adaptive Variational Monte Carlo approach developed by 
Maddox and Bittner.\cite{maddox:6465}  Here, we assume we
can write the trial density as a 
sum over $N$  Gaussian approximate functions
\begin{eqnarray}
\rho_{T}(x) = \sum_{n} G_{n}(x,{\bf c}_{n}). \label{approx}
\end{eqnarray}
parameterized by their amplitude, center, and width.    
\begin{eqnarray}
G_{n}(x,{\bf c}_{n}\}) = c_{no} e^{-c_{n2}(x-c_{n3})^{2}}
\end{eqnarray}
This trial density then is used to compute the energy
\begin{eqnarray}
E[\rho_{T}] = Tr[\rho_{T}E(x)] = \langle V_{1}\rangle + \langle Q[\rho_{T}]\rangle 
\end{eqnarray}
where $ Q[\rho_{T}] $ is the Bohm quantum potential,
\begin{eqnarray}
Q[\rho_{T} ] = -\frac{\hbar^{2}}{2m}\frac{1}{\sqrt{\rho_{T}}}\nabla^{2}\sqrt{\rho_{T}}.
\end{eqnarray}
  The energy average 
is computed by sampling $\rho_{T}(x)$ over a set of trial points $\{x_{i}\}$ and then 
moving the trial points along the conjugate gradient of 
\begin{eqnarray}
E(x) = V_{1}(x) + Q[\rho_{T}](x).
\end{eqnarray}
After each conjugate gradient step, a new set of $\bf {c}_{n}$
coefficients are determined according to an expectation maximization
criteria such that the new trial density provides the best
$N$-Gaussian approximation to the actual probability distribution
function sampled by the new set of trial points.  The procedure is
repeated until $\delta \langle E\rangle = 0$. In doing so, we
simultaneously minimize the energy and optimize the trial function.
Since the ground state is assumed to be node-less, we will not
encounter the singularities and numerical instabilities associated
with other Bohmian equations of motion based approaches.
\cite{Bohm52a,Holland93,Lopreore99,Bittner00a,Wyatt01,maddox:6465}
Moreover, the approach has been extended to very high-dimensions and
to finite temperature by Derrickson and Bittner in their studies of
the structure and thermodynamics of rare gas clusters with up to 130
atoms.  \cite{Derrickson:2006,Derrickson:2007jo}

\section{Test case: tunneling in a double well potential}
As a non-trivial test case, consider the tunneling of a particle
between two minima of a symmetric double potential well.  One can
estimate the tunneling splitting using semi-classical techniques by
assuming that the ground and excited states are given by the
approximate form
\begin{eqnarray}
\psi_{\pm} = \frac{1}{\sqrt{2}}(\phi_{o}(x) \pm \phi_{o}(-x))
\end{eqnarray}
where $\phi_{o}$ is the lowest energy state in the right-hand well in the limit the wells are 
infinitely far apart.  
% From this, one can easily estimate the splitting as \cite{Landau:1974wq}
%\begin{eqnarray}
%\delta = 4 \frac{\hbar^{2}}{m} \phi_{o}(0)\phi_{o}'(0)
%\end{eqnarray}
If we assume the localized states $(\phi_{o})$ to be gaussian,  then 
\begin{eqnarray}
\psi_{\pm} \propto \frac{1}{\sqrt{2}}(e^{-\beta(x-x_{o})^{2}}\pm e^{-\beta(x+x_{o})^{2}})
\end{eqnarray}
and we can write the superpotential as 
\begin{eqnarray}
W = \sqrt{\frac{2}{m}}\hbar\beta \left(x - x_{o}\tanh(2 x x_{o}\beta) \right).
\end{eqnarray}
From this, one can easily determine both the original potential and the partner potential as
\begin{eqnarray}
V_{1,2} &=& W^{2} \pm \frac{\hbar}{\sqrt{2m}}W' \\
&=& \frac{\beta^{2}  \hbar ^2}{m} \left( 
2   (x-x_o \tanh (2 x x_o \beta ))^2 \right. \nonumber \\
&\pm&\left. (2 x_o^2   \text{sech}^2(2 x x_o \beta )-1\right)
\end{eqnarray}
While the $V_{1}$ potential has the characteristic double minima
giving rise to a tunneling doublet, the SUSY partner potential $V_{2}$
has a central dimple which in the limit of $x_{o}\rightarrow \infty$
becomes a $\delta$-function which produces an unpaired and node-less
ground state. \cite{Cooper:1995jt} Using Eq.~\ref{chargeop}, one
obtains $\psi_{1}^{(1)} = \psi_{-} \propto A^{\dagger}\psi_{o}^{(2)}$
which now has a single node at $x = 0$.

\begin{figure}[t]
\subfigure[]{\includegraphics[width=0.45\columnwidth]{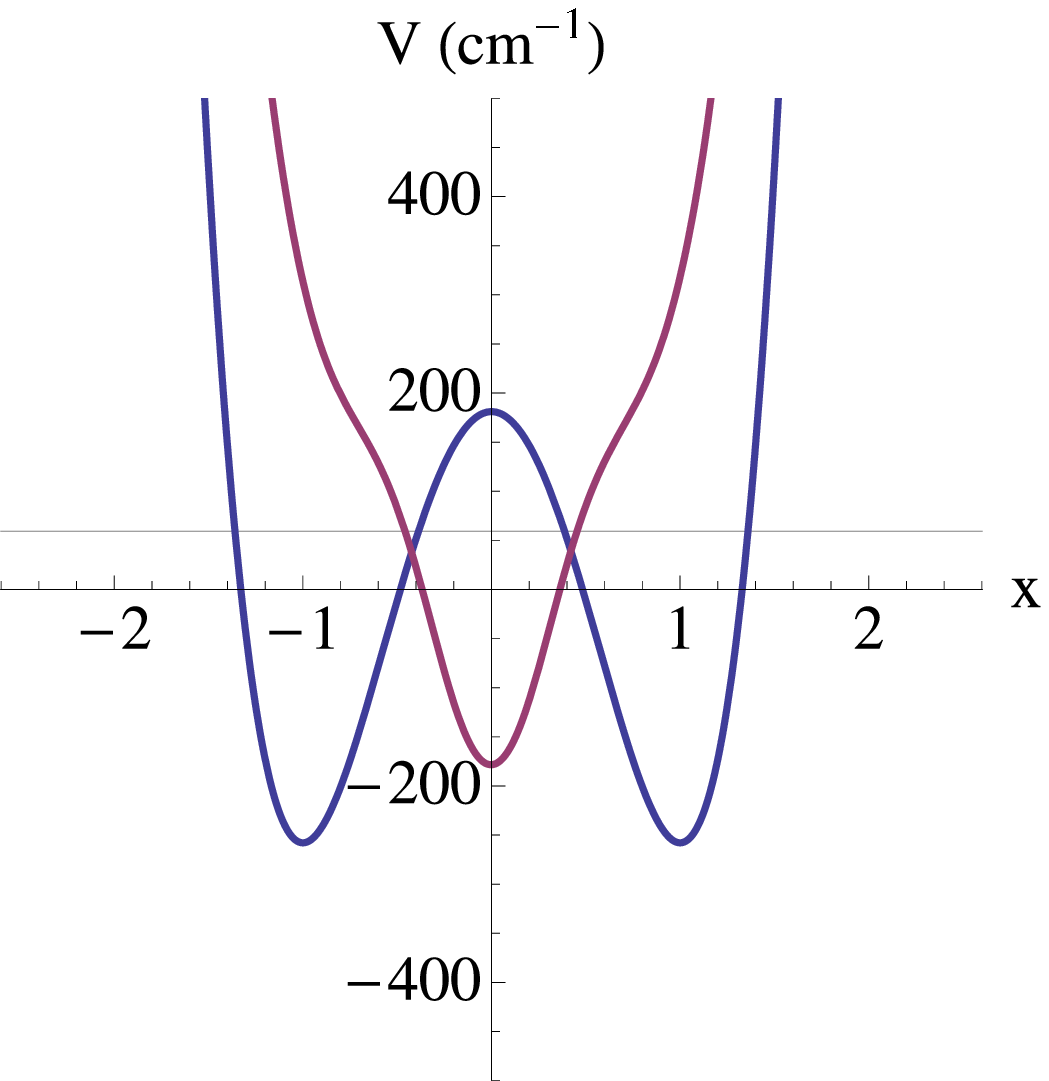}}
\subfigure[]{\includegraphics[width=0.45\columnwidth]{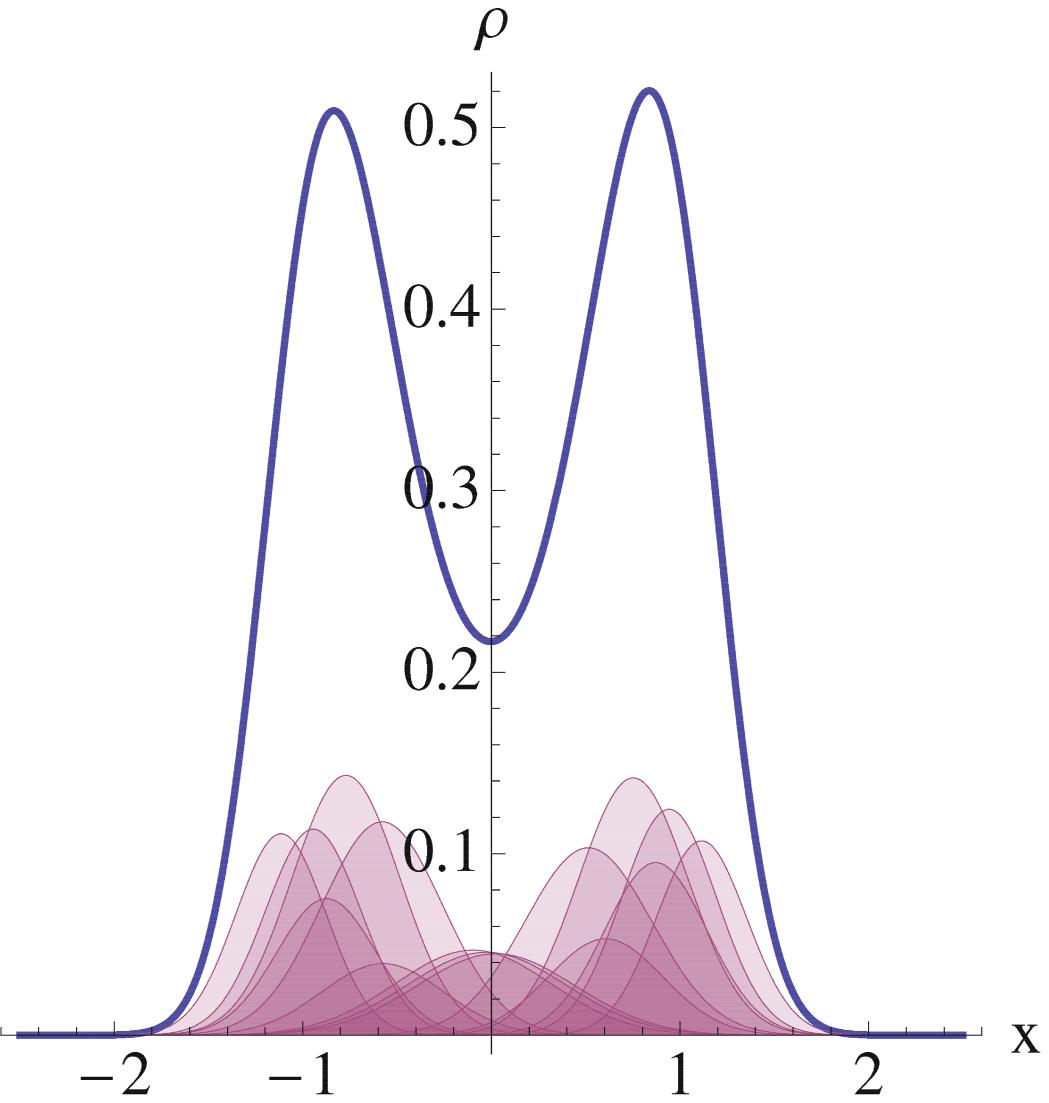}}
\caption{(a) Model double well potential(blue) and partner potential
  (purple). The energies of the tunneling doublets are indicated by
  the horizontal lines at $ V = 0\, {\rm cm}^{-1}$ and $V =59.32\,
  {\rm cm}^{-1}$ indicating the positions of the sub-barrier tunneling
  doublet.  (b) Final ground state density (blue) superimposed over
  the Gaussians used in its expansion. (purple)}\label{fig1}
\end{figure}

For a computational example,  we take the double well potential to be of the form
\begin{eqnarray}
V_{1}(x) = a x^{4} + bx^{2} + E_{o}.
\end{eqnarray}
with $a = 438.9 {\rm cm}^{-1}/(bohr^{2})$, $b = 877.8 {\rm
  cm}^{-1}/(bohr)^{4}$, and $E_{o} = -181.1 {\rm cm}^{-1}$ which (for
$m = m_{H}$ ) gives rise to exactly two states at below the barrier
separating the two minima with a tunneling splitting of 59.32 ${\rm
  cm}^{-1}$ as computed using a discrete variable representation (DVR)
approach.\cite{lig85:1400} For the calculations reported here, we used
$n_{p}=1000$ sample points and $N =15$ Gaussians and in the expansion
of $\rho_{T}(x)$ to converge the ground state.  This converged the
ground state to $1:10^{-8}$ in terms of the energy.  This, in itself
is encouraging since the accuracy of typical Monte Carlo evaluation
would be $1/\sqrt{n_{p}} \approx 3\%$ in terms of the energy.  Plots
of $V_{1}$ and the converged ground state is shown in Fig.~\ref{fig1}.
 
The partner potential $V_{2} = W^{2} + \hbar W'/\sqrt{2m}$, can be
constructed once we know the superpotential, $W(x)$.  Here, we require
an accurate evaluation of the ground state density and its first two
log-derivatives.  The advantage of our computational scheme is that
one can evaluate these analytically for a given set of coefficients.
In Fig.~\ref{fig1}a we show the partner potential derived from the
ground-state density.  Where as the original $V_{1}$ potential
exhibits the double well structure with minima near $x_{o} = \pm 1$ ,
the $V_{2}$ partner potential has a pronounced dip about $x=0$.
Consequently, its ground-state should have a simple ``gaussian''-like
form peaked about the origin.

\begin{figure}[t]
\subfigure[]{\includegraphics[width=0.45\columnwidth]{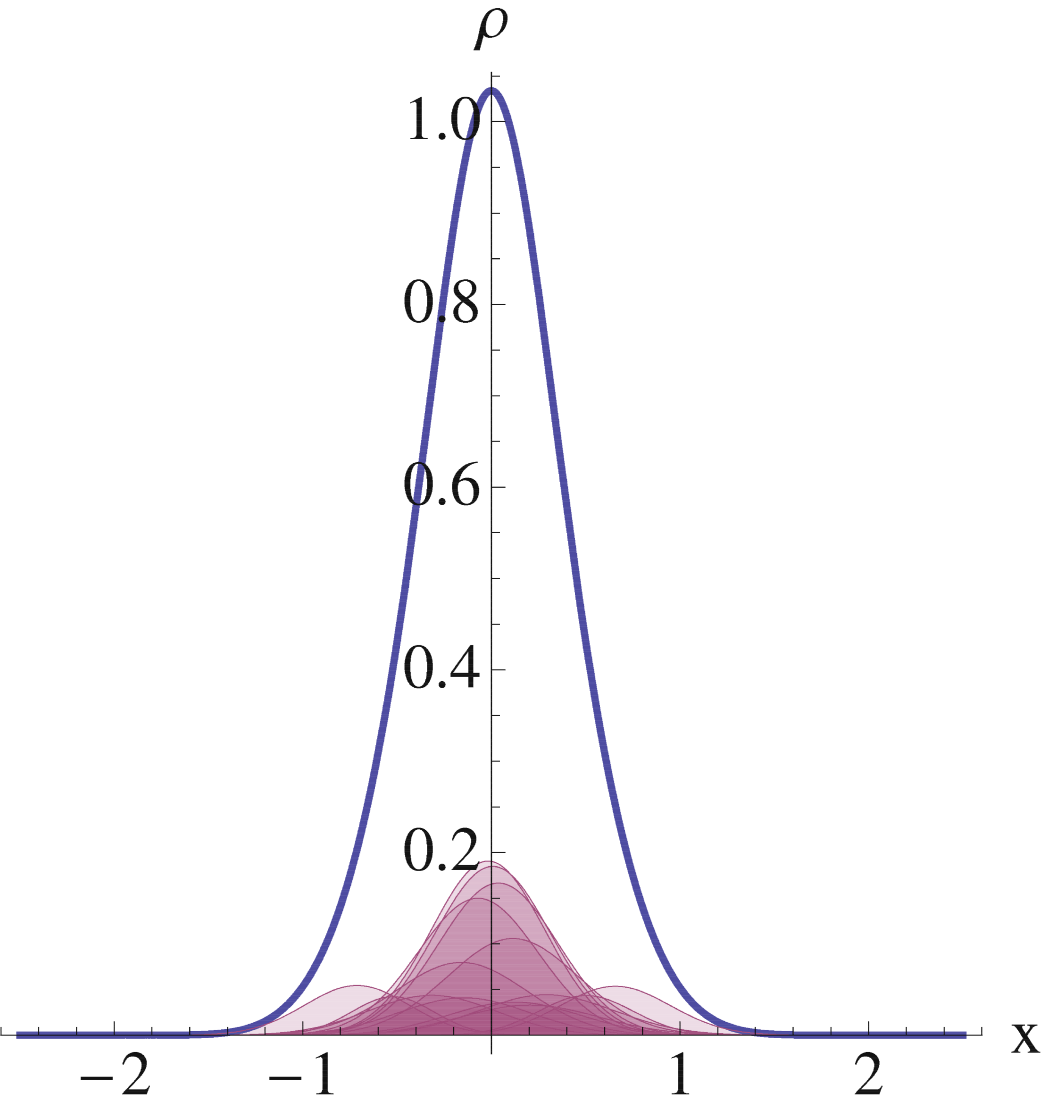}}
\subfigure[]{\includegraphics[width=0.45\columnwidth]{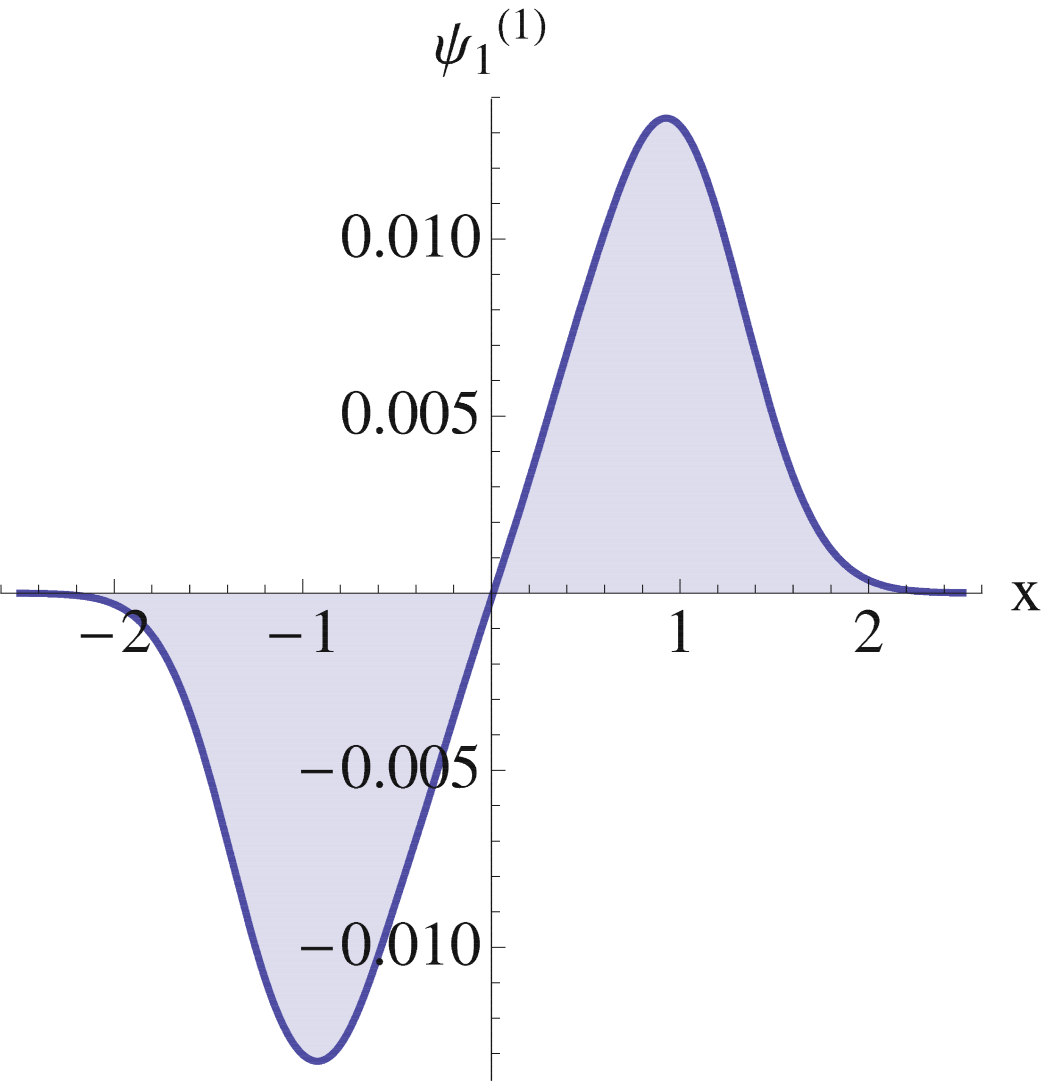}}
\caption{(a) Ground state density of the partner Hamiltonian $H_{2}$ (blue) superimposed over its
individual Gaussian components.  (b) 
Excited state  $\psi_{1}^{(1)}$ derived from the ground state of the partner potential, $\psi_{o}^{(2)}$.}\label{fig4}
\end{figure}

Once we determined an accurate representation of the partner
potential, it is now a trivial matter to re-introduce the partner
potential into the optimization routes.  The ground state converges
easily and is shown in Fig.~\ref{fig4}a along with its
gaussians. After 1000 CG steps, the converged energy is within 0.1\%
of the exact tunneling splitting for this model system.  Again, this
is an order of magnitude better than the $1/\sqrt{n_{p}}$ error
associated with a simple Monte Carlo sampling.  Furthermore,
Fig.~\ref{fig4}b shows $\psi_{1}^{(1)}\propto
A^{\dagger}\psi_{0}^{(2)}$ computed using the converged
$\rho_{0}^{(2)}$ density.  As anticipated, it shows the proper
symmetry and nodal position.

\begin{figure}[t]
%{\includegraphics[width=0.45\columnwidth]{Fig3a}}
{\includegraphics[width=\columnwidth]{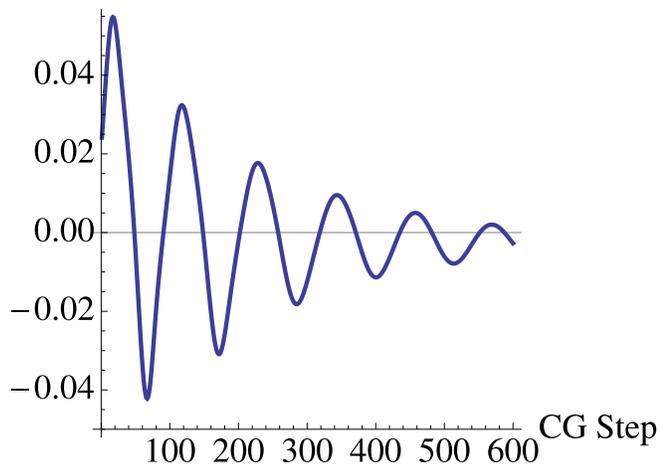}}
\caption{
Location of excited state node for the last 600 CG steps. }\label{fig3}
\end{figure}

By symmetry, one expects the node to lie precisely at the origin.
However, since we have not imposed any symmetry restriction or bias on
our numerical method, the position of the node provides a sensitive
test of the convergence of the trial density for $\rho_{0}^{(2)}$. In
the example shown in Fig.\ref{fig3}, the location of the node
oscillates about the origin and appears to converge exponentially with
number of CG steps.  This is remarkably good considering that this is
ultimately determined by the quality of the 3rd and 4th derivatives of
$\rho_{o}^{(1)}$ since these appear when computing the conjugate
gradient of $V_{2}$.  We have tested this approach on a number of
other one-dimensional bound-state problems with similar success.

\section{Extension to higher dimensions}

Having demonstrated that the SUSY approach can be used to compute
excitation energies and wave functions starting from a Monte Carlo
approach, the immediate next step is to extend this to arbitrarily
higher dimensions.  To move beyond one dimensional SUSY, Ioffe and
coworkers have explored the use of higher-order charge operators
\cite{A.A.Andrianov:1993bv,Andrianov:1995ve,0305-4470-35-6-305,Andrianov:2002gf},
and Kravchenko has explored the use of Clifford
algebras\cite{Kravchenko}.  Unfortunately, this is difficult to do in
general.  The reason being that the Riccati factorization of the
one-dimensional Schr\"odinger equation does not extend easily to
higher dimensions.  One remedy is write the charge operators as
vectors $\vec{q} = (+\vec\partial + \vec W)$ and with $\vec{q}^{+} =
(-\vec\partial + \vec{W})^{T}$ as the adjoint charge operator.  The
original Schr\"odinger operator is then constructed as an
inner-product
\begin{eqnarray}
H_{1} = \vec{q}^{+} \cdot \vec{q} .
\end{eqnarray}
Working through the vector product produces the Schr\"odinger equation
\begin{eqnarray}
H_{1}\phi = (-\nabla^{2} + W^{2}  - (\vec{\nabla}\cdot \vec{W})) \phi  = 0
\label{2dfactr}
\end{eqnarray}
and a Riccati equation of the form
\begin{eqnarray}
U(x) = W^{2} - {\rm div} \vec{W}.
\end{eqnarray}
For a 2d harmonic oscillator, we would obtain a vector superpotential of the form
\begin{eqnarray}
\vec{W} =-\frac{1}{\psi_{0}^{(1)}}{\rm grad}\psi_{0}^{(1)} =  \left(x, y\right)  = (W_{x},W_{y})
\end{eqnarray}
Let us look at the ${\rm div}\vec{W}$ part closer:
If we us the form that $\vec{W} = - \vec{\nabla}\ln\psi$, then
$-\vec{\nabla}\cdot\vec{\nabla}\ln\psi = -\nabla^{2}\ln\psi$ which for the 
2D oscillator results in ${\rm div}\vec{W} = 2$.
Thus, 
\begin{eqnarray}
W^{2} - {\rm div}\vec{W } = (x^{2} + y^{2}) - 2 
\end{eqnarray}
which agrees with the original symmetric harmonic potential.
Now, we  write the scaled partner potential as
\begin{eqnarray}
U_{2}=W^{2} +  {\rm div}\vec{W } =(x^{2} + y^{2})  + 2.\label{u2}
\end{eqnarray}
This is equivalent to the original potential shifted  by a constant amount.
\begin{eqnarray}
U_{2} = U_{1} + 4.
\end{eqnarray}
The ground state in this potential would be have the same energy as
the states of the original potential with quantum numbers $n + m = 2$.
Consequently, even with the this na\"ive factorization, one can in
principle obtain excitation energies for higher dimensional systems,
but there is no assurance that one can reproduce the entire spectrum
of states.

The problem is neither Hamiltonian $H_{2}$ nor its associated
potential $U_{2}$ is given correctly by the form implied by
Eq.~\ref{2dfactr} and Eq. \ref{u2}.  Rather, the correct approach is
to write the $H_{2}$ Hamiltonian as a {\em tensor} by taking the outer
product of the charges $\overline{H}_{2} = \vec{q} \vec{q}^{+}$ rather
than as a scaler $\vec{q} \cdot\vec{q}^{+}$.  At first this seems
unwieldy and unlikely to lead anywhere since the wave function
solutions in the second sector are now vectors rather than scalers.
However, rather than adding an undue complexity to the problem, it
actually simplifies matters considerably.  As we demonstrate in a
forthcoming paper, this tensor factorization preserves the SUSY
algebraic structure and produces excitation energies for any
dimensional SUSY system.  Moreover, this produces a scaler$\mapsto$
tensor $\mapsto$ scaler hierarchy as one moves to higher
excitations.\cite{Kouri09}

\section{Discussion}
In brief, we have used the ideas of SUSY quantum mechanics to obtain
excitation energies and excited state wavefunctions within the context
of a Monte Carlo scheme.  This was accomplished without pre-specifying
the location of nodes or restriction to a specific symmetry.  While it
is clear that one could continue to determine the complete spectrum of
$H_{1}$, the real challenge is to extend this technique to higher
dimensions.  Furthermore, the extension to multi-Fermion systems may
be accomplished through the use of the Gaussian Monte Carlo method in
which any quantum state can be expressed as a real probability
distribution. \cite{Corney,Assaad} We offer this paper as the starting
point for stimulating interest in developing numerical techniques
based upon SUSY quantum mechanics.

\begin{acknowledgments}
This work was supported in part by the National Science Foundation
(ERB: CHE-0712981) and the Robert A. Welch foundation (ERB: E-1337,
DJK: E-0608).  The authors also acknowledge Prof. M. Ioffe for
comments regarding the extension to higher dimensions.
\end{acknowledgments}

\end{document}